\documentclass[]{spie}  

\usepackage{amsmath,amsfonts,amssymb}
\usepackage{graphicx}
\usepackage[colorlinks=true, allcolors=blue]{hyperref}
\usepackage{textgreek}
\usepackage{dcolumn}
\usepackage{url}
\begin{document} 
\title{A Route to Large-Scale Ultra-Low Noise Detector Arrays for Far-Infrared Space Applications}

\author[a]{David J. Goldie}
\author[a]{Stafford Withington}
\author[a]{Christopher N. Thomas}
\author[b]{Peter A. R. Ade}
\author[b]{Rashmi V. Sudiwala}
\affil[a]{Quantum Sensors Group, University of Cambridge, Cavendish Laboratory, JJ Thomson Av., Cambridge, CB3 0HE, United Kingdom}
\affil[b]{Astronomy Instrumentation Group, Cardiff University, Cardiff, CF24 3YB, United Kingdom}

\authorinfo{Further author information: \\ E-mail: goldie@mrao.cam.ac.uk, Telephone: +44 (0)1223 337366}

\pagestyle{empty} 
\setcounter{page}{301} 

\maketitle

\begin{abstract}
%
Far-infrared detectors for future cooled space telescopes  require ultra-sensitive detectors with optical noise equivalent powers of order 
$0.2 \,\,\rm{aW/\sqrt{Hz}}$. This performance has already been demonstrated in  arrays of transition edge sensors. A critical step is demonstrating a method of fabrication and assembly that maintains the performance but that is extendable to create large-scale arrays suitable, for example, for application in  dispersive spectrometers where it may be advantageous to fabricate the array from smaller sub-arrays. Critical here are the methods of assembly and metrology that maintain the required tolerances on the spatial alignment of the  components in order to maintain overall performance. These are discussed and demonstrated. 
\end{abstract}


\section{INTRODUCTION}
\label{sec:intro}  
Far-infrared detectors for future space missions with cooled optics require ultra-sensitive detectors with technologically challenging mechanical configurations. Operation with a  grating spectrometer  implies a dimensionally large array in the spectral direction with varying spectral pixel size and spectrally-offset spatial sampling to ensure spectral coverage and increase  mapping speed.\cite{Willem_2021}
Moreover, the  sampling needs to be continuous across the overall array. Ultra-low noise transition edge sensors (TESs) with optical absorbers with backshorts and  high through-put optical feedhorns satisfy the sensitivity requirements. However, fabrication and assembly of the  detectors, backshorts and horns within the required assembly tolerances  represents  a significant challenge. 

In this paper we describe the design, fabrication, assembly, metrology and preliminary testing of arrays of ultra-low noise transition edge sensors (TESs) with optical absorbers and optical backshort arrays.
The detectors were designed with characteristics tailored for operation on the Far-Infrared Imaging Spectrometer on SPICA (SAFARI) with frequency division multiplexing (FDM) readout
 as demonstrated by SRON.\cite{Roelfsema_2018_SPICA,Wang_SRON_FDM, Audley_2020}
The overall SAFARI instrument was designed to cover the wavelength range $34-230\,\,\rm{ \mu m}$,
with 3600 TES detectors, divided  into three bands  and separate grating modules labelled (S)hort, (M)edium and (L)ong. To achieve the required sensitivities the arrays would have been cooled to 50 mK, giving optical noise equivalent power (NEP)  
of  $0.2 \,\,\rm{aW/\sqrt{Hz}}$ or better.\cite{ Audley_Safari_system_2018} 
 We have already demonstrated the core detector technology with the required characteristics across the whole of the $34-230\,\,\rm{ \mu m}$ wavelength range, \cite{Goldie_SPIE_2016,Williams_2020}
%

The design of the grating modules for an instrument such as  SAFARI requires that each band be broken-down into sub-bands 
providing contiguous sampling in both the spectral and spatial directions.
For example, for  SAFARI sub-band S1 ($34 - 43.6\,\,\rm{\mu m}$)
 an array of $90\times 5$ pixels  in the spectral-spatial
directions is required. Pixel spacings are  $P_{\rm{spectral}} = 0.8\,\,\rm{mm}$ spectrally, and $P_{\rm{spatial}}=9.775\,\,\rm{mm}$ spatially (increasing for longer wavelengths). The overall array size required for S1 is at least
 $\sim 55\times 75\,\,\rm{mm}$ depending on the wiring connection strategy to the FDM filter chips, with dimensions again increasing for the longer wavelength  bands. Moreover adjacent spatial rows needed to be offset by $P_{\rm{spatial}}/2$. In order to accommodate fabrication with typical available wafer dimensions and detector fabrication tooling, and  the capability for fabrication at the longer wavelengths, the chosen approach reported here was to break-down the arrays into sub-arrays. This introduced a particular problem. Using thin-film amorphous silicon nitride (SiN) for thermal isolation the required NEPs can only be achieved using long $\mathcal{O}(1-1.5\,\,\rm{mm})$ thermal isolation features. 
 In order to accommodate this the TES device wells must be made very large necessitating removal of a significant volume of the substrate Si.
 The remaining the Si support structure of the sub-array might become too mechanically  fragile for assembly and subsequent multiple thermal cycling.This was amongst the challenges addressed here. Likewise the spacing between the optical absorbers and both  the optical backshorts  and the horn exit apertures  needs to be precisely maintained, so that the overall flatness of the assembled sub-arrays needs to be guaranteed.
     \begin{figure} [ht]
   \begin{center}
   \begin{tabular}{c} 
   \includegraphics[width=11cm]{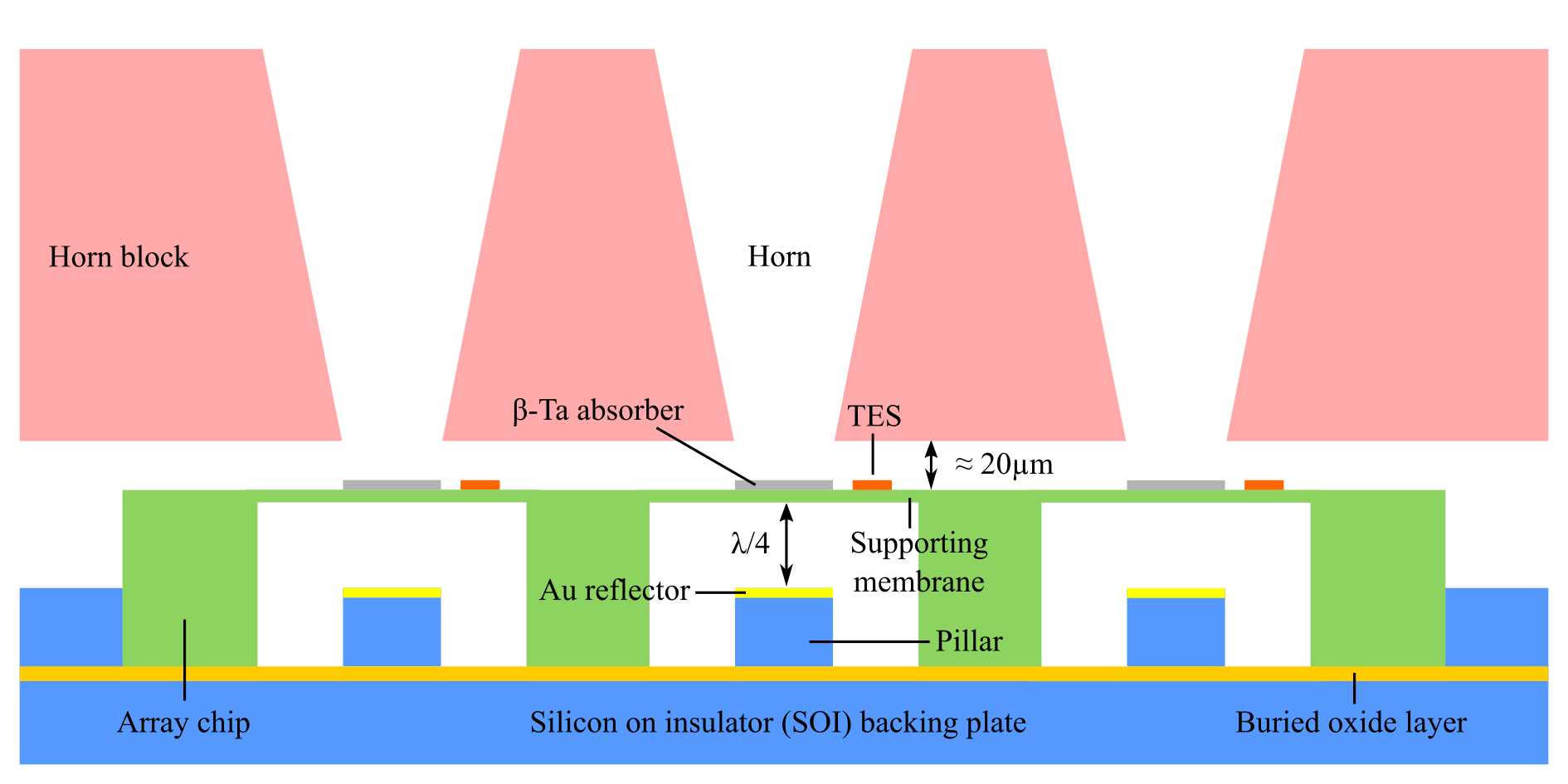}
   \end{tabular}
   \end{center}
   \caption[example] 
   { \label{fig:array_cross_section} 
Schematic cross-section (not to-scale) of the array assembled onto a backing plate with feedhorns.  }
   \end{figure} 
   Figure~\ref{fig:array_cross_section} shows a schematic cross-section of a section of an assembled  TES array, with backing plate with backshorts, and  feedhorn array. Components are indicated. The backshort to horn exit-aperture spacing can be changed for different wavelengths with an appropriate choice of the backing plate device layer thickness.

Due to the small dimensions involved we have previously found that S-band arrays were the most challenging from both the TES array fabrication and feedhorn performance points of view. The work described here describes the  fabrication, assembly, metrology and testing of small-scale sub-arrays into contiguous arrays, demonstrating array stitching, with array dimensions characteristic of those required for SAFARI at sub-band S1. The assembled detectors  were characterized  using non-multiplexed readout. 
 In order to achieve TES characteristics suitable for operation with the SRON FDM  (see Sec.~\ref{sec:FDM requirements}),  a change was required from the MoAu TES technology, previously used by the Cambridge Group to fabricate low temperature TES detectors, to higher resistance TiAu TESs. 
High performance feedhorns were also  fabricated and tested as part of the programme of work; a description of that work will be published separately. 
Although the overall programme of work was performed within the context of preparation for the SAFARI instrument, the work described here is relevant to any ultra-sensitive detector for far-infrared detection on a space satellite.

Section~\ref{sec:TES design}
describes the TES array and optical backshort array (backing plate) designs, TES and backing plate fabrication and mechanical trials showing successful array stitching. 
Section~\ref{sec:Assembly and metrology} describes array assembly into a test unit with readout and an integrated magnetic field coil suitable for assisting with TES biassing, and metrology including results of an investigation of an improved method for array mechanical clamping. 
Section~\ref{sec:Array Performance} describes preliminary measurements of the array characteristics. 
Section~\ref{sec:Conclusions} summarises the work.


\section{TES ARRAY AND BACKING PLATE DESIGN AND FABRICATION}
\label{sec:TES design}
\subsubsection{TES design for FDM compatibility}
\label{sec:FDM requirements}
The operating temperature $T$ of a voltage-biased  TES is stabilised  by electrothermal feedback (ETF) so that it operates within its  superconducting-normal-state (S-N) resistive transition,  provided the steepness of the S-N transition $T/R(dR/dT)$ is sufficiently large  ($R$  is the TES resistance, and $T$ the TES temperature). ETF means that the operating temperature  $T\simeq T_c$. ETF also  reduces the effective detector response time $\tau_\text{eff}$ and  maximises the noise contribution  of the thermal fluctuations  compared to unavoidable contributions such as those from  the readout electronics. 
The minimum NEP is achieved when noise in the heat link between $T_c$ and $T_b$ is dominant and is given by $\rm{NEP}_{phonon}=\sqrt{4k_b\gamma T_c^2 G_b}$ where $\gamma$ is a factor that depends on the temperature
 difference $\Delta T = T_c-T_b$. Modelling suggests that   $0.7<\gamma \leq1 $ for $ =0.5<\Delta T/T_c\leq 1 $.\cite{Mather}
  $G_b$ is the thermal conductance between the TES and the heat bath and is determined in our design by long, thin silicon nitride beams or legs. $G_b$ can be parmeterized such that $G_b=nK_bT_c^{(n-1)}$ where $K_b$ and $n$ depend on the characteristics of the thermal isolation material (or method) and paramaterize the power flow to the heat bath $P_b= K_b(T_c^n-T_b^n)$. We expect $1.5<n<2$ for amorphous SiN at $100<T_c<150\,\,\rm{mK}$.
 The precise $T_c$ requirement depends on the realised $K_b$ and $n$  although $T_c\sim 2\times T_b$ is sufficient for  design purposes here. 
For a satellite-mounted  instrument $T_b$ is determined by the available cooling technology that can operate in low gravity and a well-designed adiabatic demagnetization refrigerator (ADR)  suitable for space operation gives
$T_b=50\,\,\rm{mK}$.
For a target NEP of $0.2 \,\,\rm{aW/\sqrt{Hz}}$ with $T_c\sim 105\,\,\rm{mK}$, the target $G_b\sim 70\,\,\rm{fW/K}$. We expect a TES saturation power $P_{sat}\sim2-4\,\,\rm{fW}$ for expected SiN thermal parameters.

FDM read-out imposes its own requirements for the TES operating resistance and for this work a target operating resistance of $R_0\sim 30\,\,\rm{m\Omega}$
was chosen. This can been seen as a compromise between maximizing the TES current noise determined by the phonon fluctuation noise associated with $G_b$ and obtaining a sufficiently fast electrical time constant $\tau_{el}=L/R_0$, for operation with FDM and ensuring electrical stability. For the SRON FDM system the $L-C$ filter chip inductance is $L=3\,\,\rm{\mu H}$ ($C$ is the capacitance of a channel of the readout). If $\tau_{el}\le 0.1\,\,\rm{ms}$, $R_0=30\,\,\rm{m\Omega}$. For a typical S-N transition the optimum operating point (maximising the ETF effect) is achieved for $R_0\sim 0.2 R_n$ indicating a target normal state resistance $R_n=150\pm20\,\,\rm{m\Omega}$. 
Table~\ref{tab:TES Requirements} summarises TES requirements for this work. 
\begin{table}[ht]
\caption{TES Requirements} 
\label{tab:TES Requirements}
\begin{center}       
\begin{tabular}{|c|c|} 
\hline
\rule[-1ex]{0pt}{3.5ex}  \bf{Parameter}& \bf{Value}  \\
\hline
\rule[-1ex]{0pt}{3.5ex}  $T_c$ & 100-110\,$\rm{mK}$   \\
\hline
\rule[-1ex]{0pt}{3.5ex}  $G_b$ & $70\,\,\rm{fW/K}$   \\
\hline
\rule[-1ex]{0pt}{3.5ex}  $R_n$ & $150\pm20\,\,\rm{m\Omega}$    \\
\hline 
\end{tabular}
\end{center}
\end{table}

\subsection{TES Array Design}
    \begin{figure} [ht]
   \begin{center}
   \begin{tabular}{c} 
   \includegraphics[height=6cm]{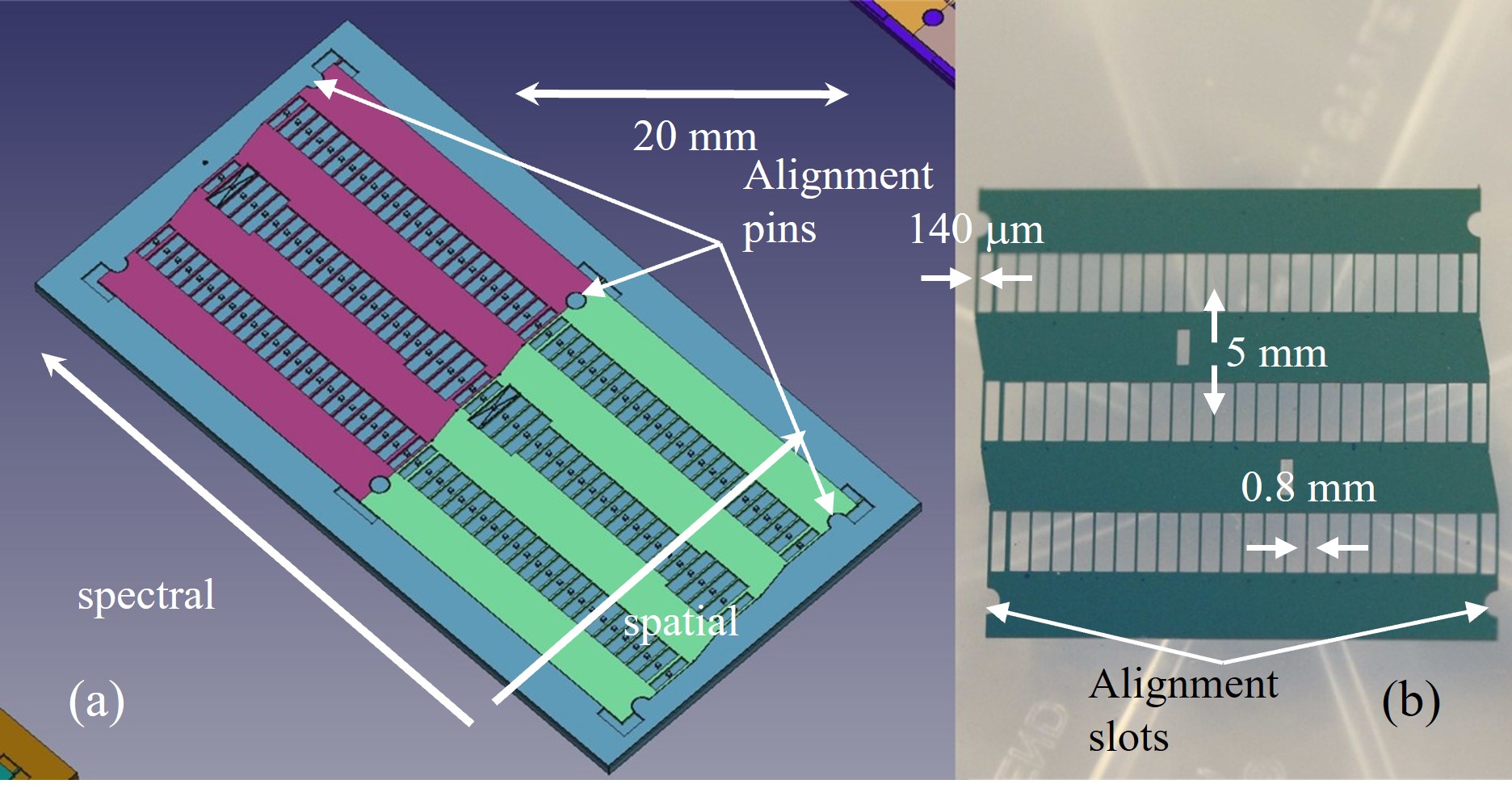}
   \end{tabular}
   \end{center}
   \caption[example] 
   { \label{fig:Fig1} 
(a)  Array and backing plate design showing two  TES sub-arrays (dark red and green colours) mounted on a backing plate (shown in blue). Spectral and spatial directions are indicated.  (b) Trial mechanical test sub-array after DRIE to shape the array chip and thermal isolation structures of the TES.}
   \end{figure} 

The  TESs detectors  were designed to be fabricated as a $60\times20\,\,\rm{\mu m}$  rectangle (i.e. 3 squares  - see Fig.~\ref{fig:TES montage}~(d)) so that using a 
TiAu bilayer with $R_{sq}=50\,\,\rm{m\Omega}$ achieved the target $R_n$. Thermal isolation was provided by four etched silicon nitride beams $1.5\,\,\rm{\mu m}$ wide and lengths in the range $1.1-1.4\,\,\rm{mm}$. The chosen pixel pitch was
 $P_{\rm{spectral}}=0.8\,\,\rm{mm}$ and  $P_{\rm{spatial}}=5\,\,\rm{mm}$. Interstitial pixels were included between the spatial rows to allow monitoring of stray-light in the final arrays. 
 
The TES sensor itself was designed to be closely thermally coupled to a superconducting Ta optical power absorber with normal-state sheet resistance
$R_{sq}\sim 380\,\,\rm{\Omega}$ chosen to be closely matched to the impedance of free-space. 
The detector (i.e TES/absorber combination) was thermally isolated from the Si substrate by photo-lithographic processing, reactive ion etching (RIE) and deep reactive ion etching (DRIE) of the Si substrate so that the detector is formed on an ``island'' that is thermally isolated from the substrate Si by thin ($\sim 1.5\,\,\rm{\mu m}$), long ($\sim 1\,\,\rm{mm}$) beams or ``legs''. For these preliminary measurements a target $T_c$ of $130\pm5\,\,\rm{mK}$  was chosen i.e. slightly higher than the instrument requirement, to allow straightforward 
measurement of optical performance using a temperature regulated ADR operating at $80\,\,\rm{mK}$. Based on the calibration of Fig.~\ref{fig:TiAu cal}, a  TiAu layup of $40/240\,\,\rm{ nm}$ was used. 
\subsection{TES Array Fabrication}
\label{sec:Array Fab}
Detectors were fabricated on Si substrates $225\,\,\rm{\mu m}$-thick with a thin $200\,\,\rm{nm}$-thick silicon nitride overlayer, and $50\,\,\rm{nm}$ $\rm{SiO}_{\rm{2}}$ DRIE etch stop. 
The front SiN layer was etched by reactive ion etching (RIE) to define the island that supports the nominally $5\,\,\rm{nm}$-thick $\beta$-phase Ta absorber ($T_c=840\,\,\rm{mK}$) and TES. 
The TESs  consisted of a TiAu bilayer with thicknesses $d_{Ti}=40\,\,\rm{nm}$, $d_{Au}=240\,\,\rm{nm}$. 
Superconducting Nb wiring connected to the TES and provided pads at the chip edges for electrical connection with Al wirebonds. RIE was used to remove the SiN from the wafer reverse and DRIE etched the Si wafer to release the detectors. 
The DRIE also defined the outline of the final wafer chip. 
All materials were deposited by magnetron sputtering under ultra-high vacuum (UHV).  
Further details of the processing are given in Refs.~\citenum{Williams_2020} and \citenum{Dorota2012}.

Two generations of arrays were fabricated;  a mechanical trial sub-array consisting of $3\times 24 $ pixels as shown in Fig~\ref{fig:Fig1} and complete detector sub-arrays of the same overall size, but where the range of $G_b$ values was expanded after successful assembly and verification of the mechanical integrity of the  trial arrays. 
\subsection{TiAu bilayers and $T_c$  calibration}
\label{sec:TiAu calibration}
In our previous work we have fabricated MoAu TESs with tunable $T_c$  in the range $100-300\,\,\rm{mK}$ that would already have satisfied the requirement for $T_c$. For superconducting-normal metal (SN) bilayers $T_c$ can be straight-forwardly tuned by changing the relative thicknesses of the superconducting  and normal metal  layers. Using MoAu bilayers to achieve $T_c\sim 100-110\,\,\rm{mK}$  typical sheet resistance $R_{sq}$ would be in the range $10-20\,\,\rm{m\Omega}$ so that a TES with a relatively large number of squares $N_{sq}=\rm{length}/\rm{width}$ would be required. 

In order to achieve a higher $R_{sq}$ whilst maintaining a reasonable aspect ratio $\sim3-4$ for the TES for this work we developed  a TiAu bilayer TES. 
 \begin{figure} [ht]
   \begin{center}
   \begin{tabular}{c} 
   \includegraphics[height=5cm]{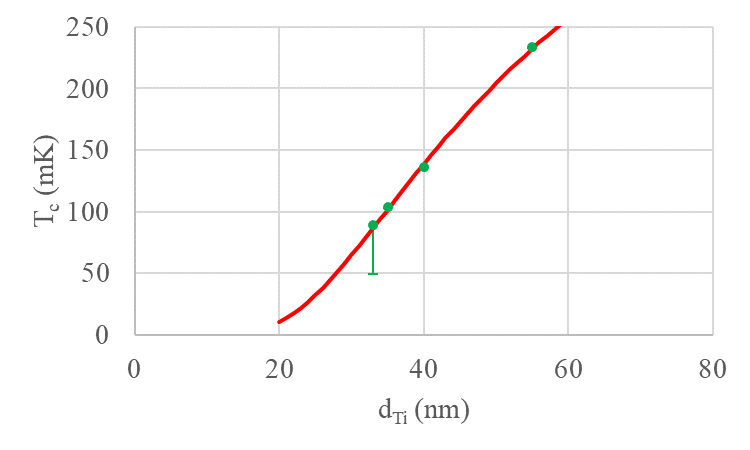}
   \end{tabular}
   \end{center}
   \caption[example] 
   { \label{fig:TiAu cal} 
Calibration trials for TiAu bilayers with Au thickness $d_{au}=240\,\,\rm{nm}$ and the Ti thickness indicated.}
   \end{figure} 
Figure~\ref{fig:TiAu cal} shows calibration data for TiAu bilayers consisting of a fixed Au thickness $d_{au}=240\,\,\rm{nm}$ and the Ti thickness indicated. The red line is a calculation based on solution of the Usadel equations to describe the superconducting proximity effect between thin SN layers.\cite{Martinis_Irwin} 
For our magnetron-sputtered Ti films we find a superconducting transition temperature $T_c=550\,\,\rm{mK}$ for films of $100\,\,\rm{nm}$ thickness, with a weak and linear dependence of $T_c$ on thickness below $100\,\,\rm{nm}$. 
In terms of the model developed in Ref~\citenum{Martinis_Irwin},
an interface transmission coefficient $t=0.07\pm0.005$ was estimated.
All films have $R_{sq}=50\pm 5\,\,\rm{m\Omega}$. 

\subsection{Backing Plate Design and Fabrication}
Backing plates were fabricated from custom-made silicon-on-insulator  (SoI) wafers.\cite{Ultrasil}
 The thickness of the SoI device layer was chosen to position the Ta absorber $10\,\,\rm{\mu m}$ above the high-reflectivity Au-coated  pillar (i.e close to $\lambda_c/4$ for sub-band S1 with $\lambda_c$ the sub-band centre wavelength) to maximize optical absorption in the free-space impedance-matched absorber for the chosen sub-band. The backing plate includes micro-machined Si alignment pins so that the precision of the assembly of sub-arrays and backing plate is determined by the DRIE. 

Figures~\ref{fig:Fig1} and \ref{fig:mechanical trial} illustrate the design and implementation of the stitching scheme. 
Figure~\ref{fig:Fig1}~(a) is  a CAD layout of two trial sub-arrays mounted on a micro-machined Si backing plate. 
 TES sub-arrays are shown in dark red and green and the backing plate is shown in light blue. 
 The backing plate forms a well into which the sub-arrays are placed. Micro-machined alignment pins on the backing plate (indicated) precisely align with alignment slots on the sub-arrays (see Fig.~\ref{fig:Fig1}~(b)). The system is straight-forwardly extendable to larger arrays. Note that the pixel offset in the spectral direction requires precise shaping of the sub-array edges. The backshort pillars and TES appear as the numerous dots in the centres of the etched Si wells. 
 Spectral and spatial directions are indicated. 
Figure ~\ref{fig:Fig1}~(b) shows a completed trial sub-array after DRIE.  Note that the Si support beams indicated on the edge of the array were designed to be reduced by the DRIE to a thickness of only $140\,\,\rm{\mu m}$ as measured.

The SoI wafers used were $100\,\,\rm{mm}$-diameter 
 with device layer thickness of $215\,\,\rm{\mu m}$ and handle Si thickness $500\,\,\rm{nm}$. This make the backing plate very robust compared to the arrays themselves.
$200\,\,\rm{nm}$-thick Au  formed the reflective backshorts. Microfabricated superconducting Nb wiring connected between the detector and the external readout circuit. All films were deposited by magnetron sputtering in UHV. 
DRIE, terminating on the buried oxide layer of the SoI wafer,  was used to define the well in which the detector array chips were positioned. The overall backing plate outline was defined by wafer dicing.
Figure~\ref{fig:array_cross_section} shows a schematic cross-section of the array assembled onto the backing plate complete with feedhorn array. 
Seating of the $225\,\,\rm{\mu m}$-thick detector array in the bachshort slot gives precise positioning of the Au reflector behind the Ta absorber. 

\subsection{Mechanical trials}
\label{sec:mechanical trials}  
     \begin{figure} [h]
   \begin{center}
   \begin{tabular}{c} 
   \includegraphics[width=16cm]{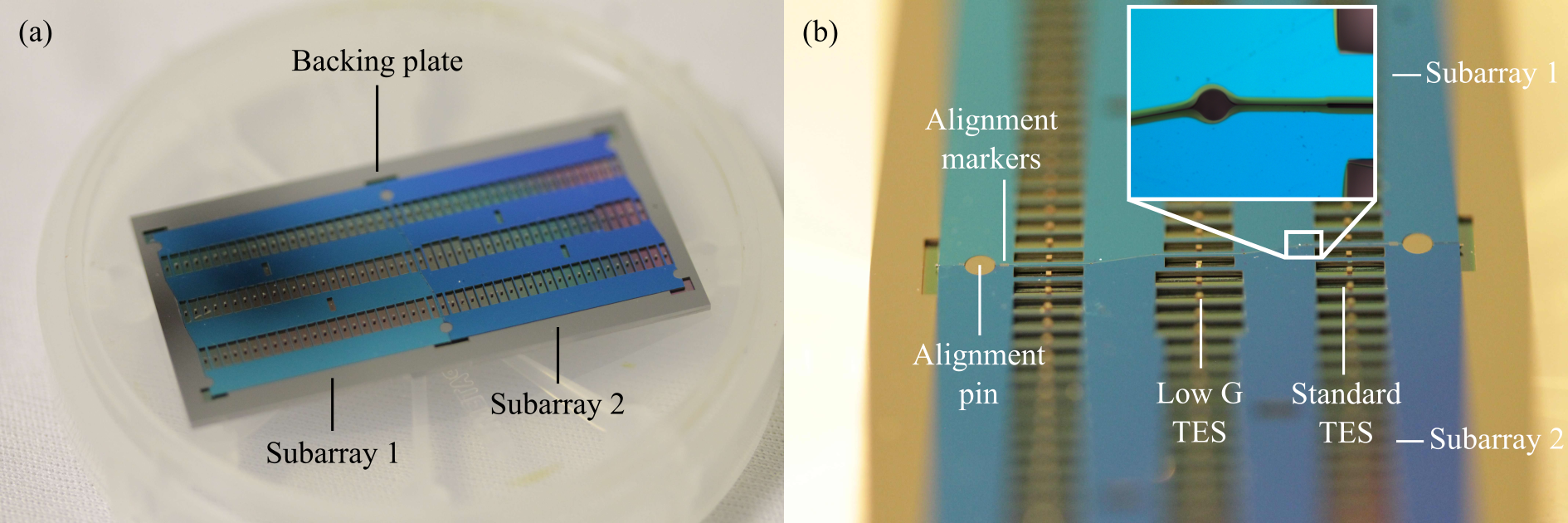}
   \end{tabular}
   \end{center}
   \caption[example] 
   { \label{fig:mechanical trial} 
(a) shows two sub-array chips assembled onto a completed backing plate. (b) shows a close-up image of the overlap region between the sub-arrays. Alignment pins fabricated by DRIE on the backing plate are indicated. The inset  shows a detail of the abutting region and includes an example of a micro-machined feature for stress relief at a Si corner. }
   \end{figure} 
Figure~\ref{fig:mechanical trial}  (a) shows two sub-array chips assembled onto a backing plate. Figure~\ref{fig:mechanical trial} (b) shows a  microscope image of the overlap region between the sub-arrays. Alignment pins fabricated by DRIE on the backing plate ensure alignment of the individual sub-arrays when assembled are indicated. The inset shows a detail of the abutting region between two sub-arrays and includes an example of a micro-machined feature for stress relief at a Si corner. The gold colour of the backing plate is an optical effect from the yellow light used for illumination.
Using lithographed alignment features fabricated on the edges of the sub-array chips, we estimated a lateral misalignment of less  than $3\,\,\rm{\mu m}$ in both spatial and spectral dimensions. 

The DRIE removes approximately $37\%$ by volume of the total Si of the sub-array chip. 
Achieving continuous spectral coverage between the sub-arrays means that the Si must be very narrow (here $140\,\,\rm{\mu m}$), and may be susceptible to damage.  
 Even so the sub-arrays have proven to be remarkably mechanically robust, 
with  no breakages on assembly or repeated thermal cycling to $60\,\,\rm{mK}$  having been observed at the time of writing. 
A particular advantage of our design is that the use of Si for both the arrays and backing plates 
eliminates differential thermal contraction between the two, which may have lead to stressing of the comparatively fragile arrays.

\subsection{Completed Arrays}
\label{sec:completed arrays}  
      \begin{figure} [ht]
   \begin{center}
   \begin{tabular}{c} 
   \includegraphics[width=14cm]{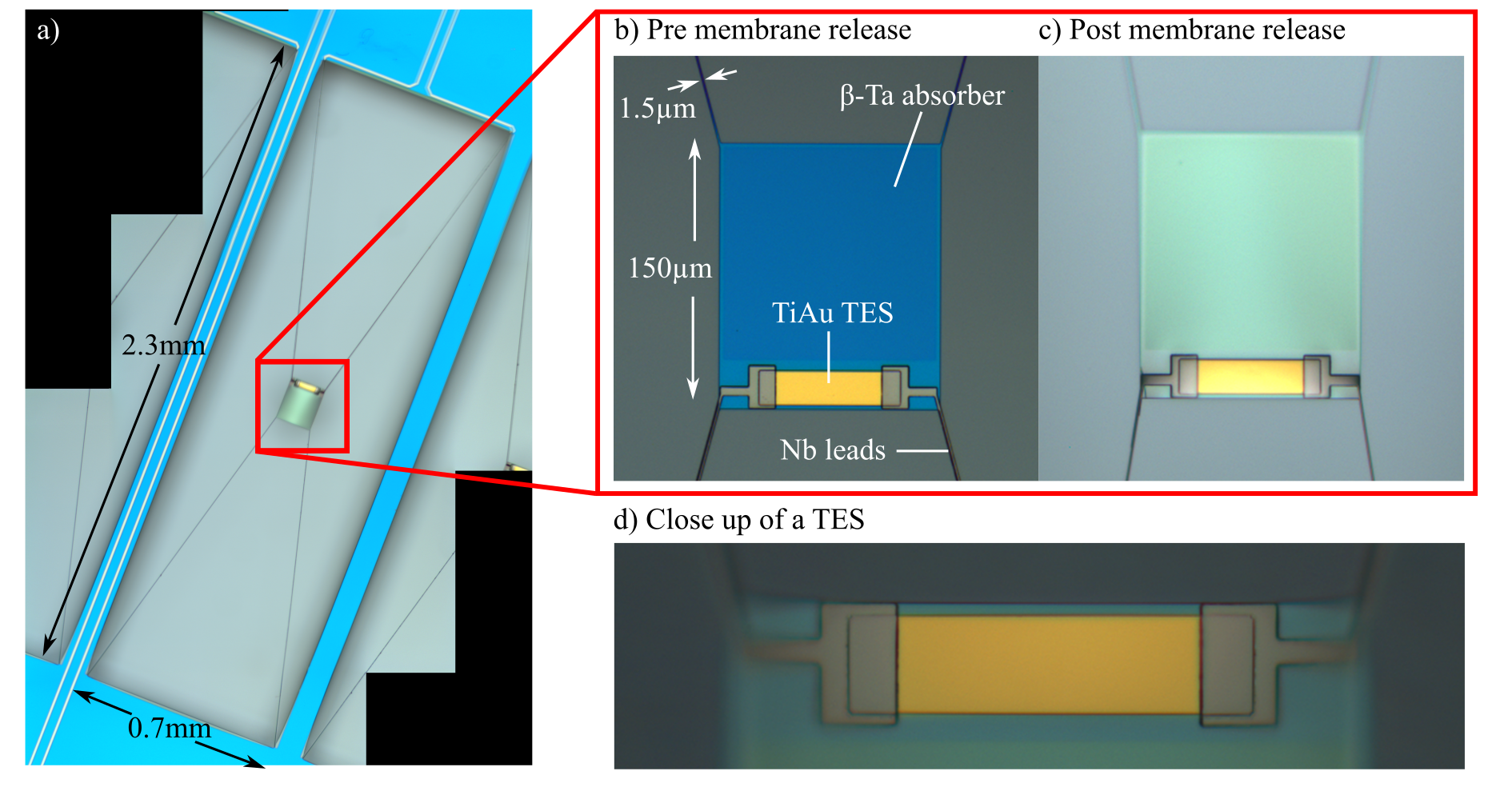}
   \end{tabular}
   \end{center}
   \caption[example] 
   { \label{fig:TES montage} 
(a) Microscope image of a single pixel after DRIE, (b) zoomed image of the TES and Ta optical absorber before and (c) after DRIE to remove the underlying Si, and (d) close-up of a TiAu TES. }
   \end{figure} 
   Figure~\ref{fig:TES montage} shows images from a completed sub-array. (a) shows an overall view of a pixel with TES and optical  absorber. The SiN support legs ($1100\times 1.5\times0.2\,\,\rm{\mu m}$) are visible as the ``X'' shaped lines connecting to the corners of the TES/absorber island. Overall dimensions of the pixel are indicated. The Si bars between pixels are $100\,\,\rm{\mu m}$ wide. 
  (b) and (c) are micrographs of the TES, absorber and Nb wiring, before and after DRIE. The $8\,\,\rm{nm}$-thick Ta absorber causes the colour change in the central area of (c). (d) shows a close up of the three square TiAu TES with Nb contacts. 
   
\section{ASSEMBLY AND METROLOGY}
\label{sec:Assembly and metrology}
\subsection{Assembly}
   \begin{figure} [ht]
   \begin{center}
   \begin{tabular}{c} 
   \includegraphics[width=14cm]{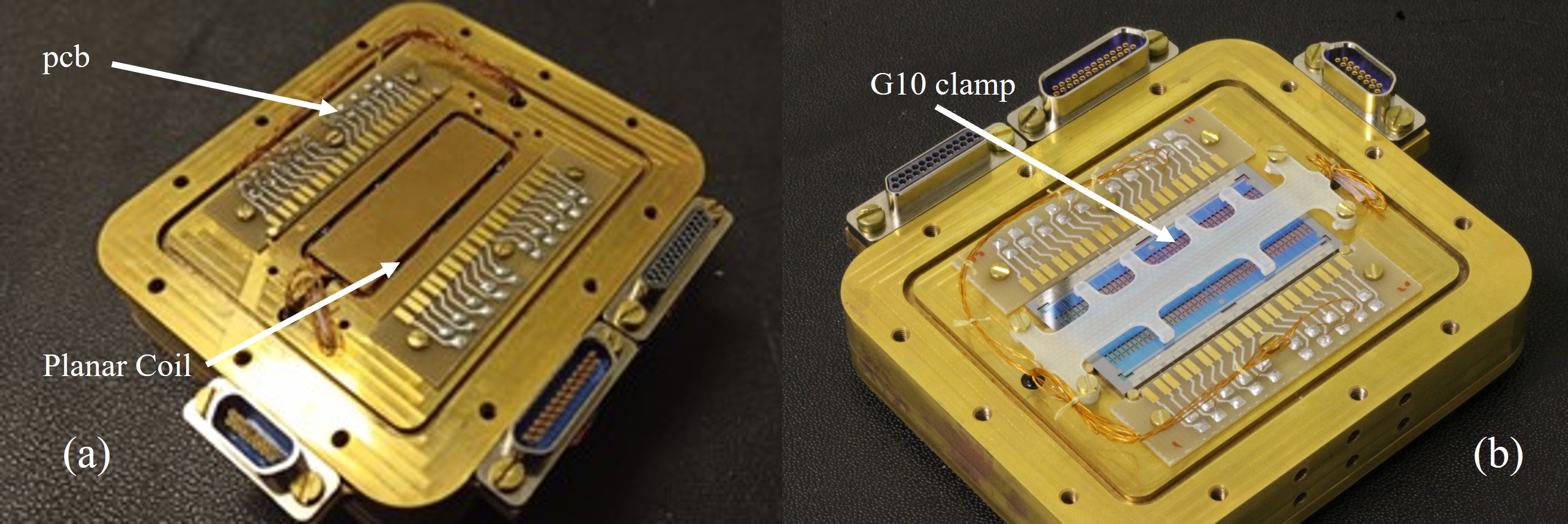}
   \end{tabular}
   \end{center}
   \caption[example] 
   { \label{fig:DC unit assy} 
(a) Image of the part-assembled test unit. The planar superconducting field coil is indicated. (b) Complete assembly before installation of the light-tight lid. }
   \end{figure}     
   For cold testing the sub-arrays were mounted into a light-tight test unit. The test unit included the detectors with superconducting fan-out PCBs and, in a separate compartment directly behind the detectors,  6 channel SQUID electronics and TES bias circuitry (not shown). Electrical connections between the PCB and SQUID/bias circuitry used superconducting NbTi twisted-pairs with  light-tight feed-throughs between the two compartments. 
    The two-stage SQUIDs used for readout were  provided by Physikalisch-Technische Bundesanstalt (PTB)
 or purchased from Magnicon. \cite{Magnicon}

 The test unit also included a planar superconducting coil that was formed from $200\,\,\rm{\mu m}$ NbTi wire in a machined slot  directly behind the polished detector mounting surface. The coil allows application of small ($\mathcal{O}(100\,\,\rm{\mu T}))$ magnetic field $B$
with low applied currents that are sufficient to significantly reduce the TES critical current $J_c$  and hence  facilitate driving the TES  its superconducting to normal state for detector operation. This feature would be particularly relevant to FDM readout, especially on a satellite, where TES bias circuits may be unable to provide the necessary current to switch-on the TES. For the measurements reported here the coil was not essential, but we found a reduction in  $J_c$  of order a factor 10 with an applied magnetic field close to the expected first minimum of the  Fraunhofer $J_c-B$ diffraction pattern of the TES when treated as an $\rm{S-S^{\prime}}$ superconducting weak-link. Such a coil would  be extremely useful in a space application not only for biassing the array but also, for example, for cancelling stray fields, or multiple such coils might have advantages for biasing or field-cancellation across a large array with spatially varying fields. The coil has very low mass compared to using  large external field coils.
  
Figure~\ref{fig:DC unit assy} 
shows images of the test unit unit during assembly. (a) shows the  front compartment of the test unit before installation of the detector array. The planar coil is indicated. (b) shows the assembled unit with backing plate and array. The fibre-glass G10 clamp used to clamp down the detector assembly for these preliminary tests is also shown. 
The assembled arrays were cooled in an ADR that gave a base temperature of $57\,\,\rm{mK}$. 
\subsection{Metrology}
      \begin{figure} [ht]
   \begin{center}
   \begin{tabular}{c} 
   \includegraphics[width=12cm]{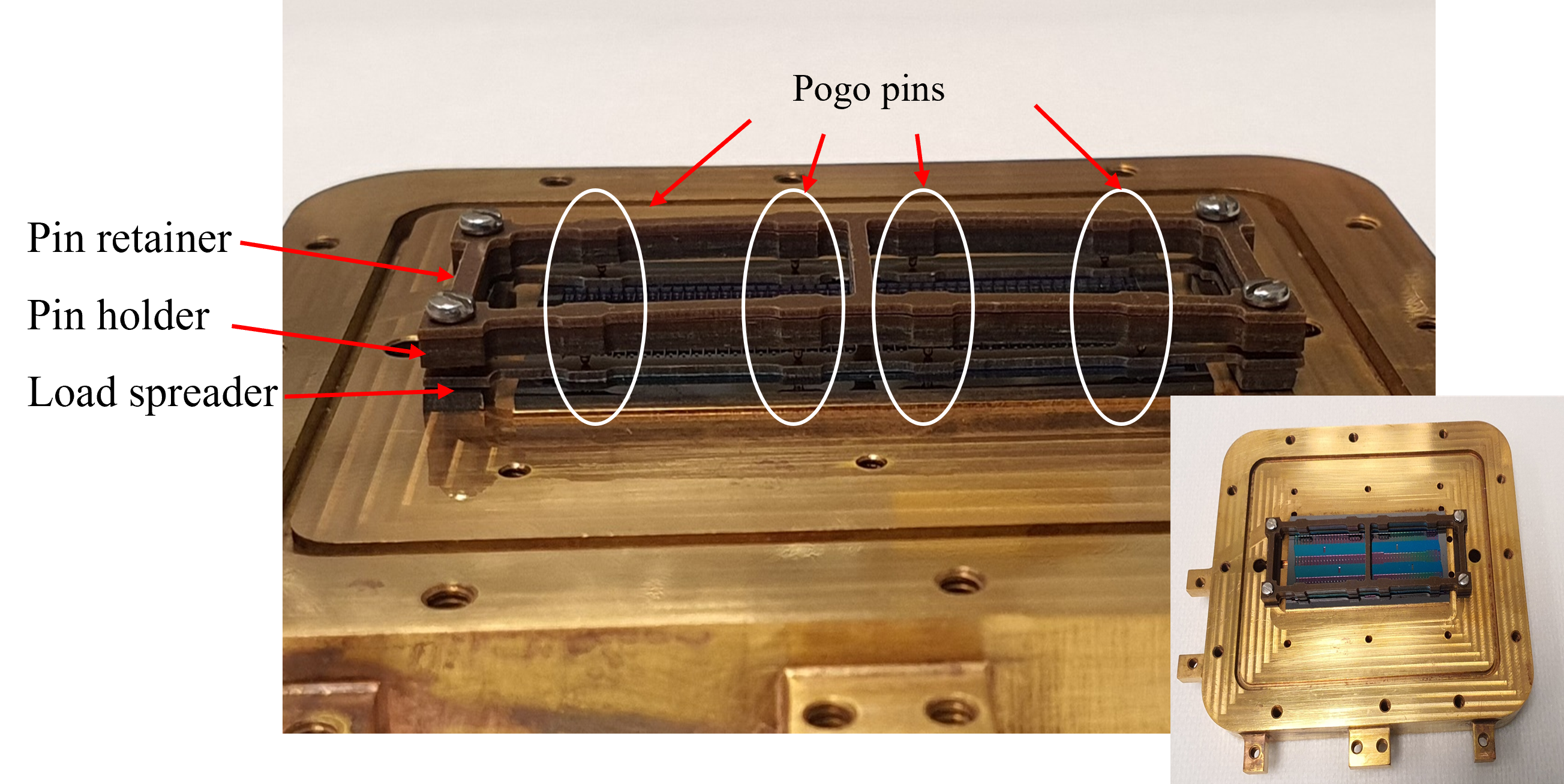}
   \end{tabular}
   \end{center}
   \caption[example] 
   { \label{fig:Clamping scheme} 
 Image of the modified array clamping scheme using pogo pins. }
   \end{figure} 
Previous measurements using a non-contact surface profiler on a smaller S-band array and a similar G10 mounting scheme
 showed wafer flatness of order $\pm 3\,\,\rm{\mu m}$ across the array.\cite{Goldie_SPIE_2016}
  Similar measurements for the mounting scheme shown in 
 Fig.~\ref{fig:DC unit assy} (b) gave a typical sub-array surface flatness of  $\pm 8\,\,\rm{\mu m}$. 
   Two causes were identified for the enhanced distortion observed. First, the sub-array chips themselves show enhanced distortion. This was measured when the chips were unclamped and results from rebalancing of internal stresses after removal of much of the Si. Second, the G10 clamp  only provides a four-point contact that exacerbates the problem and the clamp itself distorts slightly when assembled.
  Whilst not problematic for the measurements reported here, this would be a significant problem both for positioning and optical performance if combined with a horn array where the exit apertures of the horns need to placed $20\,\,\rm{\mu m}$ above the Ta absorber surface (see Fig.~\ref{fig:array_cross_section}).
\subsubsection{Modified array clamping scheme}
\begin{figure} [h]
   \begin{center}
   \begin{tabular}{c} 
   \includegraphics[width=10cm]{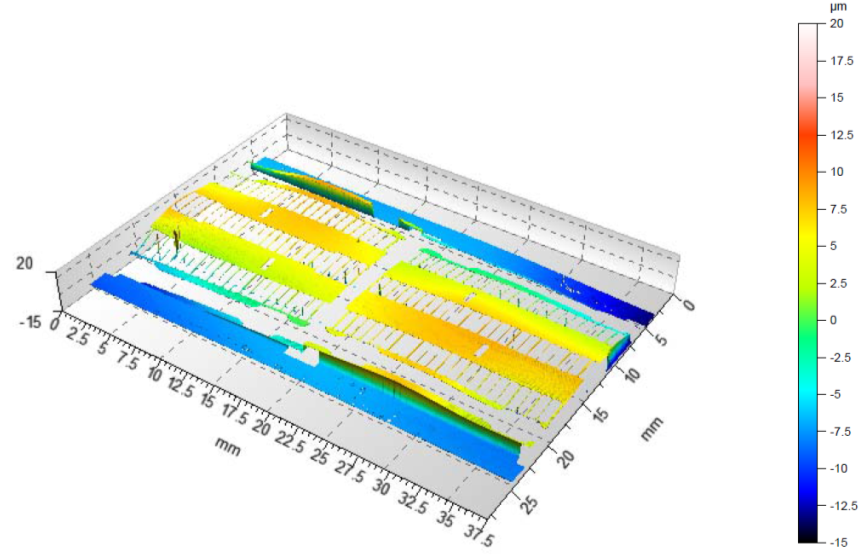}
   \end{tabular}
   \end{center}
   \caption[example] 
   { \label{fig:Polytech} 
Non-contact surface profile measurement of across the sub-arrays of Fig.~\ref{fig:Clamping scheme} using the modified clamping scheme.  }
   \end{figure} 
To reduce the distortion a prototype clamping system based on commercially available electrical spring-loaded pogo pin contacts was tested. The test clamp components, comprising a pin holder, a load spreader and top pin retainer, were machined from Tufnol Kite sheet. Figure~\ref{fig:Clamping scheme} shows an image of the modified assembly using an engineering-grade test unit. 
  
   Figure~\ref{fig:Polytech}  shows a non-contact surface profiler measurement of the flatness of the modified clamping scheme. The upper surfaces of the two sub-arrays are visible as the coloured-coded areas that show the height distributions.
 There is appreciable improvement in detector chip flatness using this system - we estimate a flatness of order $\pm5\,\,\rm{\mu m}$. The use of additional clamping points should improve the array flatness further. 
   In future iterations of the technology, the pogo pins would also be incorporated directly into the horn block for compactness.
\section{ARRAY PERFORMANCE}
\label{sec:Array Performance}
\begin{table}[ht]
\caption{TES measurements} 
\label{tab:TES measurements}
\begin{center}       
\begin{tabular}{|c|c|c|c|c|} 
\hline
\rule[-1ex]{0pt}{3.5ex}  \bf{TES} & $\mathbf{R_n\,\rm{\bf{(m\Omega)}}}$  & $\mathbf{T_c\,\rm{\bf{(mK)}}}$  & $\mathbf{G_b\,\rm{\bf{(fW/K)}}}$  & \textbf{Legs ($\boldsymbol{\mu}$m)} \\
\hline
\rule[-1ex]{0pt}{3.5ex} 1 & $170\pm3$  & $115\pm3$ &  $90\pm5$ & $1100\times1.5\times0.2$ \\
\hline
\rule[-1ex]{0pt}{3.5ex} 2 & $165\pm3$ & $128\pm3$ &  $100\pm5$  & $1100\times1.5\times0.2$\\
\hline
\rule[-1ex]{0pt}{3.5ex} 3 & $165\pm3$ & $109\pm3$  &$60\pm5$  & $1400\times1.5\times0.2$\\
\hline 
\end{tabular}
\end{center}
\end{table}
Table~\ref{tab:TES measurements} summarises preliminary TES measurements from sub-array 1. 
$R_n$, $T_c$ and $G_b$ were calculated directly from measured current-voltage characteristics or derived from the related $P_b$ as a function of $T_b$. 
Normal state resistances were close to design value. We found  $K_b=200\pm20\,\,(\rm{fW/K}^n)$ and $n=1.7\pm0.1$, close to expectations. A slightly greater range of $T_c$ values were indicated by the measurements than would be expected based on measurements of the bilayer calibration samples. However, the variations in $T_c$  provided a  greater variation in $G_b$ than would otherwise have been measured. The source of the variations in $T_c$ remains an area of investigation. 

\begin{table}[ht]
\caption{Measured TES saturation powers at $R_0=30\,\,\rm{m\Omega}$ and phonon-limited NEP assuming $\gamma=0.7$. } 
\label{tab:TES saturation}
\begin{center}       
\begin{tabular}{|c|c|c|} 
\hline
\rule[-1ex]{0pt}{3.5ex}  \bf{TES} & $\mathbf{P_{sat}\,\rm{\bf{(fW)}}}$  & \bf{NEP} $\rm{\bf{(aW/\sqrt{Hz})}}$     \\
\hline
\rule[-1ex]{0pt}{3.5ex} 1 & $2.95$  & $0.205$   \\
\hline
\rule[-1ex]{0pt}{3.5ex} 2 & $4.12$ & $0.25$  \\
\hline
\rule[-1ex]{0pt}{3.5ex} 3 & $1.88$ & $0.17$    \\
\hline 
\end{tabular}
\end{center}
\end{table}
Table~\ref{tab:TES saturation} summarises measured TES saturation powers determined at $R_0=30\,\,\rm{m\Omega}$  and the calculated phonon-limited NEP with noise-modifier $\gamma=0.7$. $P_{sat}$ values are as expected given the range of measured $T_c$ and within the range expected from the design. The calculated phonon-limited NEP brackets the NEP  required for an instrument like SAFARI. 
\section{CONCLUSIONS} 
\label{sec:Conclusions}
We have described the design, fabrication, assembly and preliminary testing of  TiAu TESs with superconducting transition temperatures of order $110\,\,\rm{mK}$ and phonon-limited NEP satisfying the requirements of an instrument like SAFARI. The TESs are thermally coupled to large area optical absorbers suitable for integration with feedhorns for future optical characterization. The arrays were formed from sub-arrays that were micro-machined by DRIE and successfully assembled into continuous detector arrays in a geometry suitable for spectral-spatial imaging without loss of spectral coverage at the sub-array boundaries. 

The assembled arrays have proved to be  mechanically robust and have been repeatedly thermally cycled from room temperature to below $60\,\,\rm{mK}$ without any observed damage or loss of individual pixels despite the inherent fragility of the sub-arrays themselves. Using lithographed alignment features fabricated on the edges of the arrays we
demonstrated  a lateral misalignment of less  than $3\,\,\rm{\mu m}$ in both spatial and spectral dimensions
 between the sub-arrays using micromachined alignment features on the sub-arrays and backing plates. 
We have developed and demonstrated an improved array clamping scheme using mechanical spring-loaded pins giving a vertical flatness of mounted sub-arrays flatness of order $\pm5\,\,\rm{\mu m}$ that is already suitable for implementation in the presence of horn arrays.

\subsection{Acknowledgments}
This work was funded by the European Space Agency under contract number 4000130201/20/NL/IB/gg.

The authors are very grateful to Gert de Lange, Jan van der Kuur, Pourya Khosropanah, Dennis van Loon  and Damian Audley for their advice and comments in order to understand the requirements imposed by the SRON FDM readout.


\end{document}